\documentclass[preprint, superscriptaddress, preprintnumbers, amsmath, amssymb]{revtex4}

\allowdisplaybreaks
\allowdisplaybreaks[4]
\usepackage{CJK}
\usepackage{graphicx}% Include figure files
\usepackage{dcolumn}% Align table columns on decimal point
\usepackage{bm}% bold math
\usepackage[dvipdfm,
            pdfstartview=FitH,
            CJKbookmarks=true,
            bookmarksnumbered=true,
            bookmarksopen=true,
            colorlinks=true,
            pdfborder=001,
            linkcolor=blue,
            anchorcolor=blue,
            citecolor=blue,
            ]{hyperref}

\begin{document}

\begin{CJK}{GBK}{song}

\title{Reexamine the nuclear chiral geometry from the orientation of the angular momentum}

\author{Q. B. Chen}
\affiliation{Physik-Department, Technische Universit\"{a}t
M\"{u}nchen, D-85747 Garching, Germany}
\author{J. Meng}\email{mengj@pku.edu.cn}
\affiliation{State Key Laboratory of Nuclear Physics and Technology,
             School of Physics, Peking University, Beijing 100871, China}%
\affiliation{Yukawa Institute for Theoretical Physics, Kyoto
             University, Kyoto 606-8502, Japan}
\affiliation{Department of Physics, University of Stellenbosch,
             Stellenbosch, South Africa}%

\date{\today}

\begin{abstract}

The paradox on the previous  interpretation for the nuclear chiral
geometry based on the effective angle has been clarified by
reexamining the system with the particle-hole
configuration $\pi (1h_{11/2})^1 \otimes \nu(1h_{11/2})^{-1}$
and rotor with deformation parameter $\gamma=30^\circ$.
It is found that the paradox is caused by the
fact that the angular momentum of the rotor is much smaller than
those of the proton and the neutron near the bandhead.
Hence, it does not support a chiral rotation interpretation
near the bandhead. The nuclear chiral geometry based on the
effective angle makes sense only when the angular momentum
of the rotor becomes comparable with those of the proton
and the neutron at the certain spin region.

\end{abstract}

\maketitle

%%%%%%%%%%%%%%%%%%%%%%%%%%%%%%%%%%%%%%%%%%%%%%%%%%%%%%%%%%
%                    begin  introduction
%%%%%%%%%%%%%%%%%%%%%%%%%%%%%%%%%%%%%%%%%%%%%%%%%%%%%%%%%%

Chirality is a topic of general interest in the sciences,
such as chemistry, biology, and physics. An object
or a system is chiral if it is not identical to its mirror
image, and cannot be superposed on its mirror image through any
combination of rotations and translations.

The phenomenon of chirality in nuclear physics was initially
introduced by Frauendorf and Meng in 1997~\cite{Frauendorf1997NPA}
for a fast rotating nucleus with triaxially deformed shape and
high-$j$ valence particle(s) and valence hole(s). In that circumstances,
the collective angular momentum is favor of aligning along
the nuclear intermediate axis that provides the largest moment of inertia, while
the angular momentum vectors of the valence particles (holes) align
along the short (long) axis. Such arrangement makes the
three angular momenta perpendicular to each other and form either a
left- or a right-handed system. Reversing the direction of the
component of the angular momentum on one of principal axes changes
the chirality of the system. This phenomenon appears in the body-fixed
reference frame where the spontaneous breaking of the chiral symmetry
happens. In the laboratory reference frame, however, due to the quantum
tunneling of the total angular momentum between the left- and
right-handed system, the broken chiral symmetry is restored. Then, the
chiral doublet bands, i.e., a pair of nearly degenerate $\Delta I=1$ bands
with the same parity, are expected to be observed~\cite{Frauendorf1997NPA}.

After the pioneering work on the chirality in nuclei~\cite{Frauendorf1997NPA},
the chiral symmetry in atomic nuclei has become one of the most intriguing phenomena
that has attracted significant attentions and intensive studies from both
experimental and theoretical sides in the last two decades. On the experimental
side, the chiral doublet bands were first observed in four $N=75$ isotones
in 2001~\cite{Starosta2001PRL}. So far, more than forty pairs of
chiral doublet bands candidates have been reported in the $A\sim 80$, 100, 130, and 190
mass regions. For recent reviews, see Refs.~\cite{J.Meng2010JPG, J.Meng2014IJMPE,
Bark2014IJMPE, J.Meng2016PS, Raduta2016PPNP, Starosta2017PS, Frauendorf2018PS}.
With the prediction~\cite{J.Meng2006PRC} and
confirmation~\cite{Ayangeakaa2013PRL} of the multiple chiral
doublets (M$\chi$D) in a single nucleus, the investigation of the chirality
continue to be one of the hottest topic in nuclear physics~\cite{J.Peng2008PRC,
J.M.Yao2009PRC, J.Li2011PRC, Droste2009EPJA, Q.B.Chen2010PRC, Hamamoto2013PRC,
Tonev2014PRL, Lieder2014PRL, Rather2014PRL, Kuti2014PRL, C.Liu2016PRL, Grodner2018PRL,
H.Zhang2016CPC, J.Li2018PRC}.

As demonstrated in Refs.~\cite{Frauendorf1997NPA, Frauendorf2001RMP}, the
chirality of nuclear rotation results from not only the static (the triaxial shape)
but also the dynamics (the angular momentum) properties of the nucleus. This is
quite different from the chirality in chemistry, which is of static nature that
characterizes just the geometrical arrangement of the atoms. Hence, it is of
importance to examine the angular momentum geometry in order to verify
whether the pair of nearly degenerated doublet bands are chiral doublet bands
or not. To achieve this goal, one can investigate: (1) the angular momentum
components of the rotor, the particle(s), and the hole(s) along
the three principal axes (e.g., in Refs.~\cite{Frauendorf1997NPA, Olbratowski2004PRL,
Olbratowski2006PRC, S.Q.Zhang2007PRC, B.Qi2009PLB, B.Qi2009PRC, Lawrie2010PLB,
Q.B.Chen2010PRC, Hamamoto2013PRC, H.Zhang2016CPC, Petrache2016PRC}); (2) the
distributions of the angular momentum components on the three intrinsic
axes ($K$ plot) (e.g., in Refs.~\cite{J.Peng2003PRC, S.Q.Zhang2007PRC, B.Qi2009PLB,
B.Qi2009PRC, Q.B.Chen2010PRC, H.Zhang2016CPC, F.Q.Chen2017PRC}); (3)
the effective angles between the angular momenta of the rotor, the particle(s),
and the hole(s) (e.g., in Refs.~\cite{Starosta2001NPA, Starosta2002PRC, S.Y.Wang2007PRC}); (4) the
orientation parameter of the system (e.g., in Refs.~\cite{Starosta2001NPA,
Starosta2002PRC, Grodner2018PRL}); (5) the distributions of the tilted angles of
the angular momentum in the intrinsic frame (azimuthal plot) (e.g., in
Ref.~\cite{F.Q.Chen2017PRC}); etc.

\begin{figure}[!ht]
  \begin{center}
    \includegraphics[width=7.5 cm]{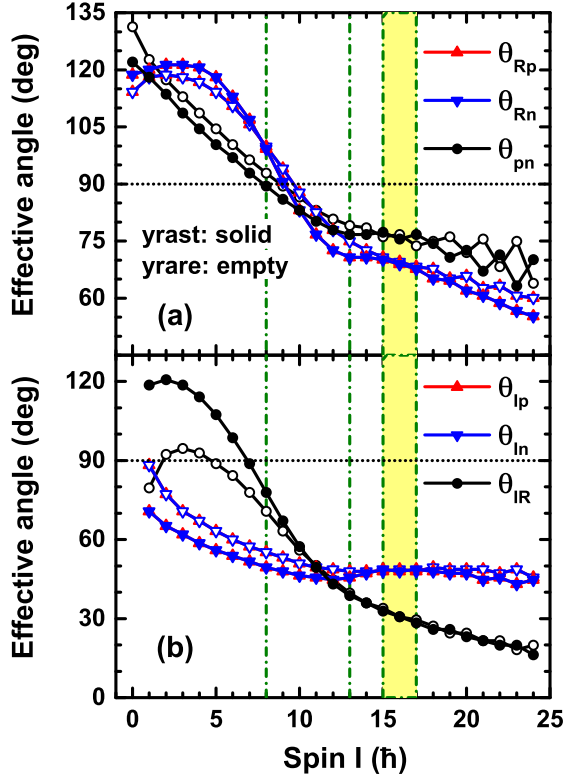}
    \caption{The effective angles $\theta_{pn}$, $\theta_{Rp}$, $\theta_{Rn}$,
    $\theta_{Ip}$, $\theta_{In}$, and $\theta_{IR}$ as functions of spin
    for the yrast and yrare bands.}\label{fig1}
  \end{center}
\end{figure}

It is known now that chiral rotation (or static chirality) can exist only above
a certain critical frequency~\cite{Olbratowski2004PRL, Olbratowski2006PRC, P.W.Zhao2017PLB,
Grodner2018PRL}. Namely, at low spin the chiral vibrations, understood as the oscillation of
the total angular momentum between the left- and the right-handed configurations
in the body-fixed frame, exists. This suggests that the orientation of the angular momenta
of the rotor, the particle(s), and the hole(s) are {\it planar} at the
bandhead of the chiral bands. However, it is noted that the effective angles between
any two of the three angular momenta are closed to $90^\circ$ as shown for
the yrast band of $^{126}$Cs~\cite{S.Y.Wang2007PRC} (see also Fig.~\ref{fig1}).
This is the paradox which motivates us to reexamine the angular momentum geometries
of the rotor, the particle(s), and the hole(s) in the chiral doublet bands.

Theoretically, various approaches have been developed extensively
to investigate the chiral doublet bands. For example, the particle rotor
model (PRM)~\cite{Frauendorf1997NPA, J.Peng2003PRC,
Koike2004PRL, S.Q.Zhang2007PRC, B.Qi2009PLB}, the titled axis
cranking (TAC) model~\cite{Dimitrov2000PRL, Olbratowski2004PRL, Olbratowski2006PRC,
P.W.Zhao2017PLB}, the TAC plus random-phase approximation (RPA)~\cite{Almehed2011PRC},
the collective Hamiltonian method~\cite{Q.B.Chen2013PRC, Q.B.Chen2016PRC}, the
interacting boson-fermion-fermion model~\cite{Brant2008PRC}, and the angular momentum
projection (AMP) method~\cite{Bhat2012PLB, Bhat2014NPA, F.Q.Chen2017PRC, Shimada2018PRC_v1}.
In this work, the PRM will be used. The basic microscopic inputs for PRM can be obtained
from the constrained covariant density functional theory (CDFT)~\cite{J.Meng2006PRC, J.Meng2016book,
Ayangeakaa2013PRL, Lieder2014PRL, Kuti2014PRL, C.Liu2016PRL, Petrache2016PRC}.
PRM is a quantal model consisting of the collective rotation and the intrinsic single-particle
motions, which describes a system in the laboratory reference frame. The total Hamiltonian
is diagonalized with total angular momentum as a good quantum number. The energy
splitting and quantum tunneling between the doublet bands can be obtained directly.
Hence, it is straightforward to be used to investigate the angular momentum geometries
of the chiral doublet bands.

The detailed formalism of PRM can be found in Refs.~\cite{Frauendorf1997NPA, J.Peng2003PRC,
Koike2004PRL, S.Q.Zhang2007PRC, B.Qi2009PLB}. In the calculations,
a system of one $h_{11/2}$ proton particle and one $h_{11/2}$ neutron
hole coupled to a triaxial rigid rotor with quadruple deformation
parameters $\beta = 0.23$ and $\gamma=30.0^\circ$ are taken as the example
to illustrate the angular momentum geometry.
In addition, the irrotational flow type of moments of inertia
$\mathcal{J}_k=\mathcal{J}_0\sin^2(\gamma-2k\pi/3)$ $(k=1, 2, 3)$
with $\mathcal{J}_0=30~\hbar^2/\textrm{MeV}$ are used.

%%%%%%%%%%%%%%%%%%%%%%%%%%%%%%%%%%%%%%%%%%%%%%%%%%%%%%%%%%
%                    begin  results and discussion
%%%%%%%%%%%%%%%%%%%%%%%%%%%%%%%%%%%%%%%%%%%%%%%%%%%%%%%%%%

The effective angle $\theta_{pn}$ between the proton ($\bm{j}_p$) and neutron
($\bm{j}_n$) angular momenta is defined as~\cite{Starosta2002PRC},
\begin{align}\label{eq1}
 \cos\theta_{pn}=\frac{\langle \bm{j}_p\cdot \bm{j}_{n}\rangle}
  {\sqrt{\langle \bm{j}_p^2 \rangle} \sqrt{\langle \bm{j}_n^2 \rangle}},
\end{align}
and similarly for $\theta_{Rp}$, $\theta_{Rn}$, $\theta_{Ip}$, $\theta_{In}$,
and $\theta_{IR}$. Here, the subscripts $p$, $n$, $R$, and $I$ denote
the proton, the neutron, the rotor, and the total spin, respectively,
and $|\rangle$ is the wave function of the yrast or yrare bands. In
geometry, any three vectors lie in a planar only when the
sum of any two angles between the vectors equals the other one or the sum of
the three angles equals $360^\circ$.

\begin{figure*}[!ht]
  \begin{center}
    \includegraphics[height=9.0 cm]{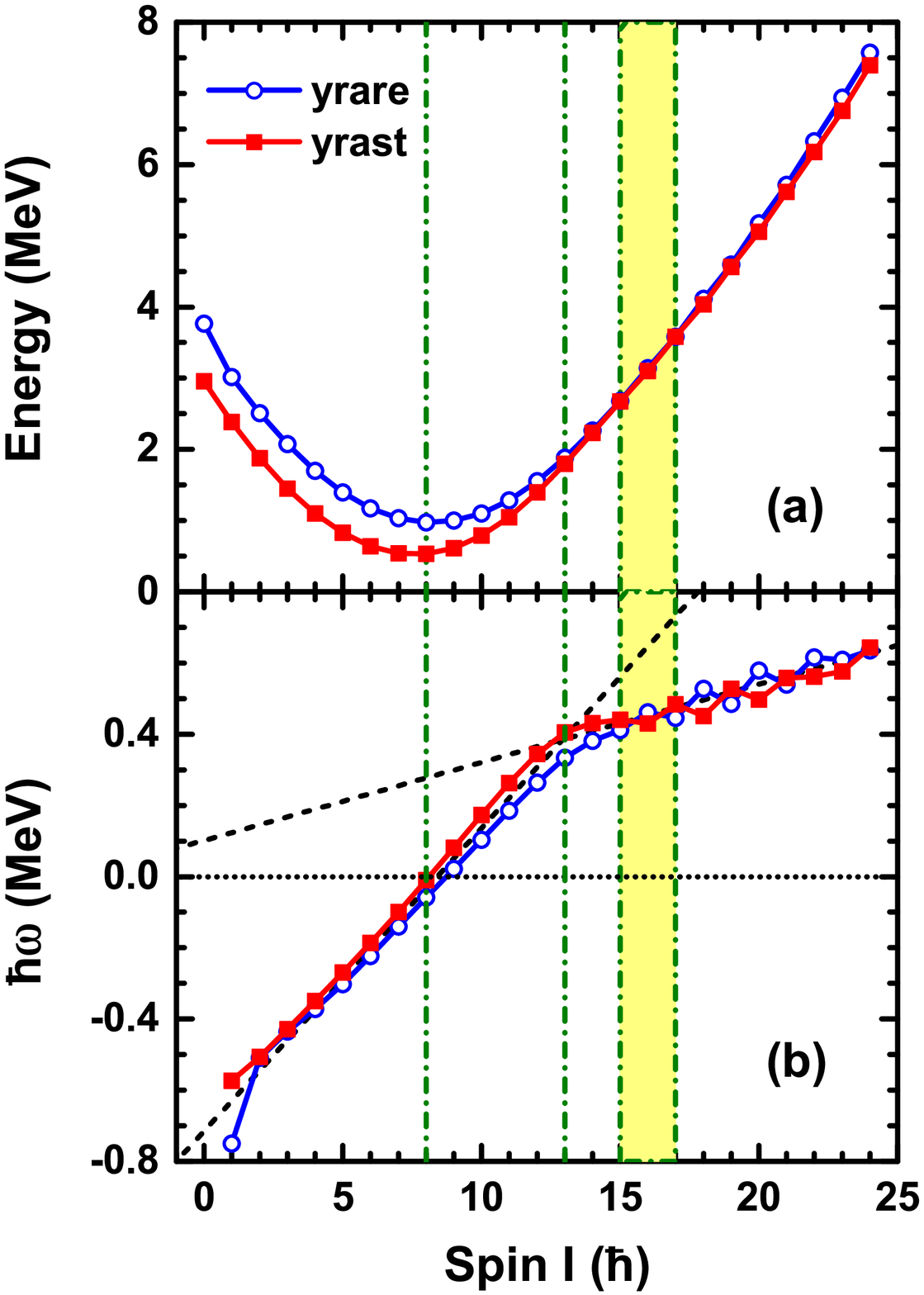}~~
    \includegraphics[height=9.0 cm]{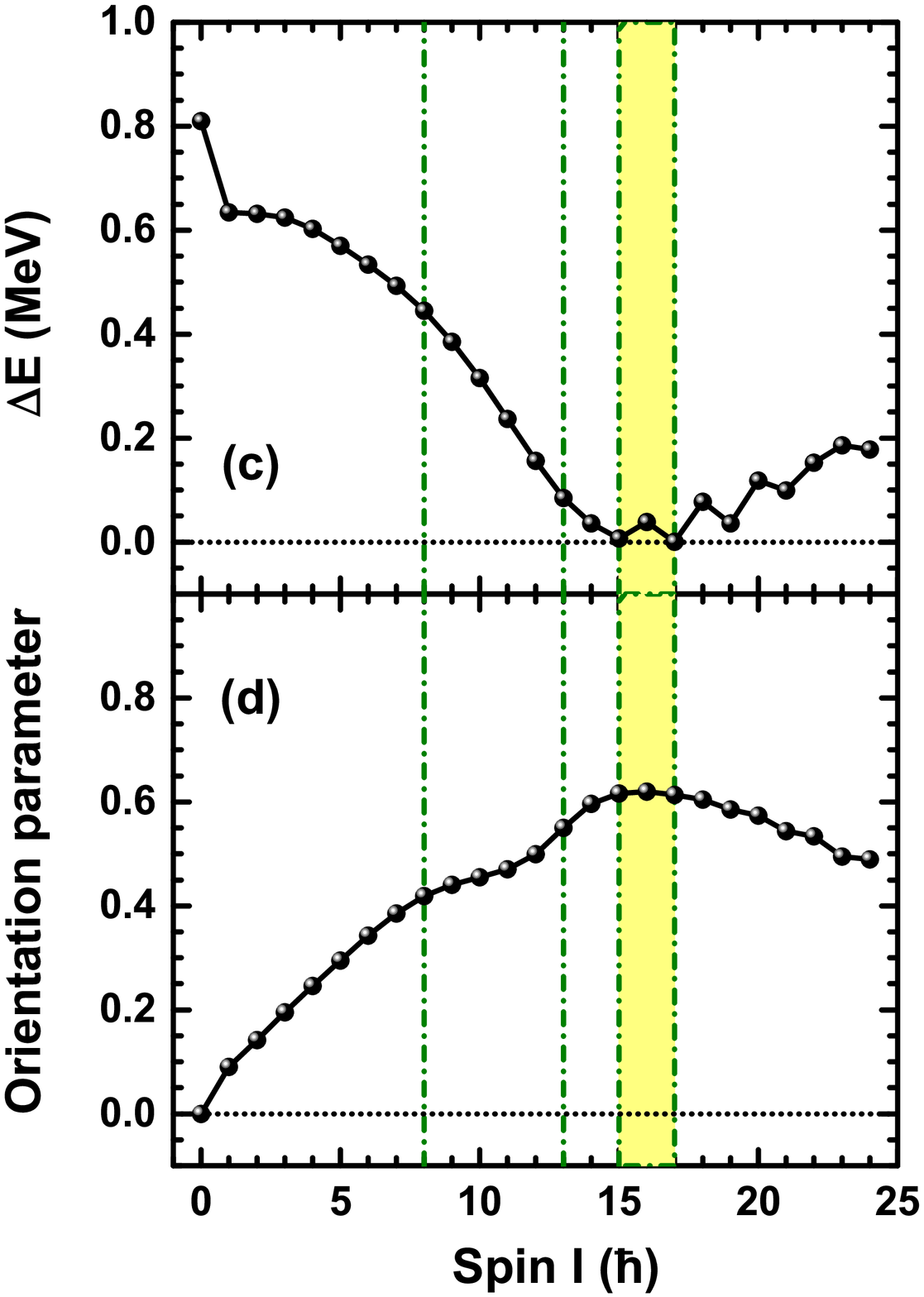}
    \caption{(a) Energy spectra as functions of spin for the yrast and yrare
    bands. (b) The extracted rotational frequencies as functions of spin.
    (c) Energy difference between the doublets. (d) The normalized
    orientation parameter calculated by Eq.~(\ref{eq6}).}\label{fig2}
  \end{center}
\end{figure*}

\begin{figure*}[!ht]
  \begin{center}
    \includegraphics[width=16.0 cm]{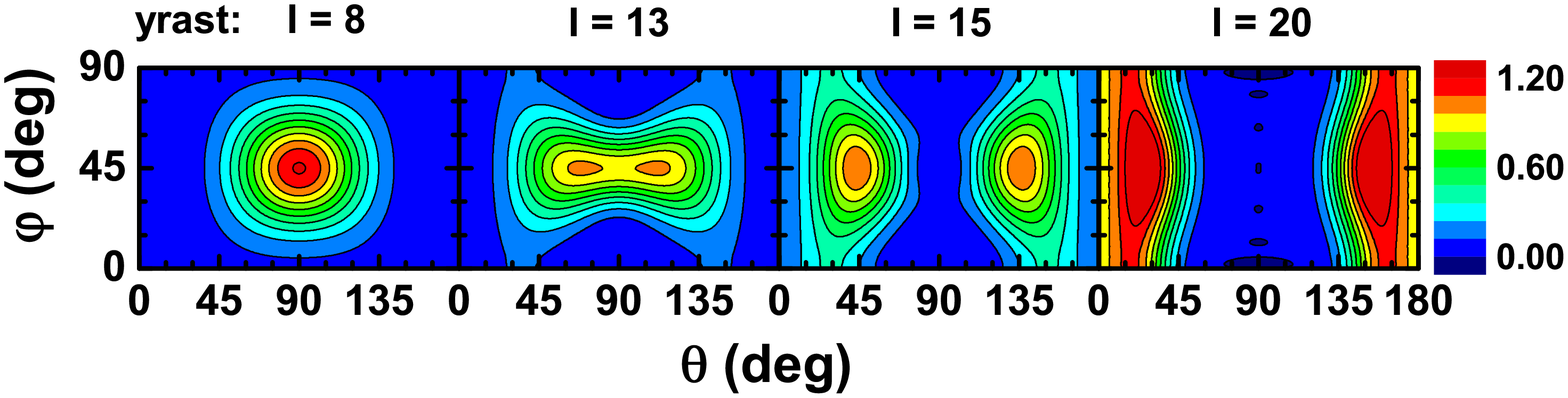}
    \includegraphics[width=16.0 cm]{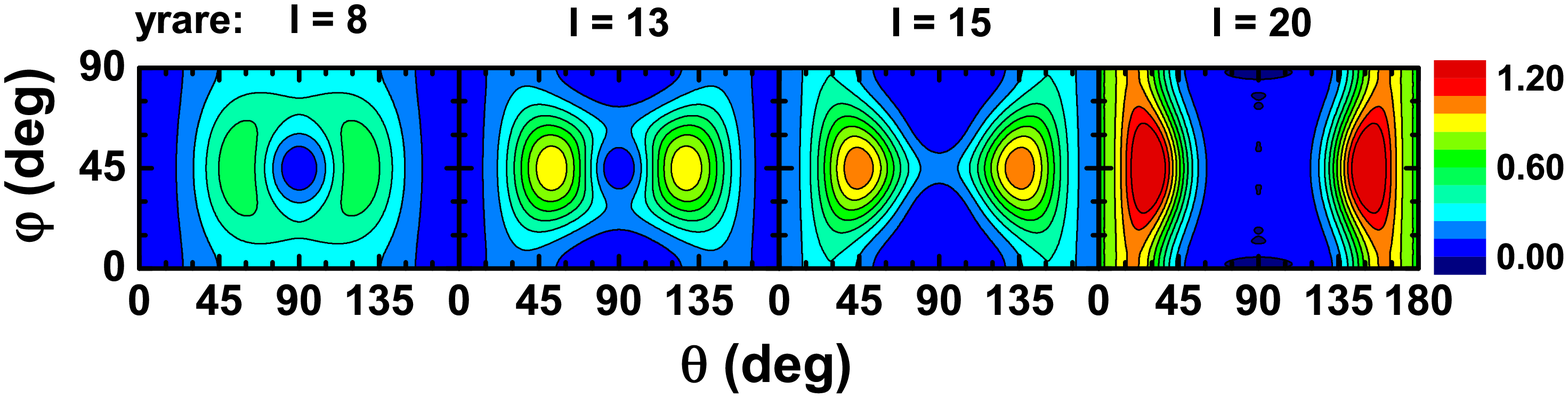}
    \caption{The azimuthal plots, i.e., profiles for the orientation of the
    angular momentum on the $(\theta,\varphi)$ plane, calculated at $I=8$,
    13, 15, and $20\hbar$, for the yrast and yrare bands.}\label{fig3}
  \end{center}
\end{figure*}

In Fig.~\ref{fig1}, the obtained effective angles $\theta_{pn}$, $\theta_{Rp}$,
$\theta_{Rn}$, $\theta_{Ip}$, $\theta_{In}$, and $\theta_{IR}$ as functions
of spin for the yrast and yrare bands are presented. The dashed-dotted lines at
$I=8$, $13$, and $15$-$17\hbar$ label the bandhead, the onset
of aplanar rotation, and the static chirality, respectively, which are
based on the Figs.~\ref{fig2} and \ref{fig3} showing later.

From Fig.~\ref{fig1}(a), it is observed that the effective
angles $\theta_{pn}$, $\theta_{Rp}$, and $\theta_{Rn}$
are about 120$^\circ$ around $I=0\hbar$, i.e., the
angular momenta $\bm{j}_p$, $\bm{j}_n$ and $\bm{R}$ have to cancel out
to obtain the total spin zero. The sum of the three effective angles
equals to $\sim 360^\circ$, i.e., the three angular
momenta indeed lie in a plane.

The three effective angles gradually decrease with spin and drop
to $\sim 90^\circ$ at the bandhead ($I=8\hbar$), which leads to the conclusion
that the angular momenta $\bm{j}_p$, $\bm{j}_n$, and $\bm{R}$ are nearly mutually
perpendicular to each other in Ref.~\cite{S.Y.Wang2007PRC}. This is the
paradox with respect to the understanding of chiral vibration near the bandhead.

At the static chiral region ($15\leq I \leq 17\hbar$), the three effective angles of
the doublet bands are rather similar. Note that the values of these three effective
angles are about $70^\circ$, a bit far from $90^\circ$. It seems that
the aplanar rotation at this spin region is less than that near
the bandhead. This is also contradiction with our empirical understanding
for the static chirality and need to be solved.

The obvious odd-even staggering behaviors of $\theta_{pn}$ at $I\geq 20\hbar$
and of $\theta_{Rp/Rn}$ at $I\geq 21\hbar$ indicate a strong signature
splitting of a principle axis rotation.

For the effective angles with respect to the total spin, $\theta_{Ip/In}$
are smaller than $90^\circ$ at the whole spin region, which implies
that $\bm{j}_p$ and $\bm{j}_{n}$ align toward the total spin.
At $I\geq 13\hbar$, they do not vary much. For $\theta_{IR}$,
it is larger than $90^\circ$ for the yrast band below the bandhead.
This means that the $\bm{R}$ anti-aligns along the total spin.
The decreasing of $\theta_{IR}$ indicates that the role of the
rotor becomes more and more essential. Meanwhile, the differences of
$\theta_{Ip/In}$/$\theta_{IR}$ between the doublet bands become
smaller with spin. At $I=15$-$17\hbar$, they are almost
the same. At the high spin region ($I>20\hbar$),
$\theta_{Ip/In}$/$\theta_{IR}$ show small staggering behaviors.

To solve this paradox, we first reexamine the energy
spectra of the chiral doublet bands in Fig.~\ref{fig2}(a).
Similar results has already been presented in
Refs.~\cite{Frauendorf1997NPA, B.Qi2009PRC, Q.B.Chen2010PRC, H.Zhang2016CPC},
but here lower spin (from $0\hbar$) ones will be focused. At
$I\leq 8\hbar$, the energies of the doublets decrease with
spin, since the collective rotations have not yet started. In the shown figures,
the dashed-dotted line at $I=8\hbar$ is plotted to label this bandhead
position. At the intermediate spin region (around $I=15$-$17\hbar$), near energy
degeneracies of doublets are found. To show this more clearly, the energy
difference between the doublet bands $\Delta E(I)=E_{\textrm{yrare}}(I)
-E_{\textrm{yrast}}(I)$ is shown in Fig.~\ref{fig2}(c). One sees that
it decreases first and then increases. At $I=15$-$17\hbar$,
it is the smallest, corresponding to the best degeneracy and
static chirality (marked by a shadow). At high spin region ($I\geq 18\hbar$),
it shows an odd-even staggering behavior, caused by the signature
splitting of the principal axis rotation.

From the energy spectra, the rotational frequencies $\hbar\omega(I)=E(I)-E(I-1)$
are extracted~\cite{Frauendorf1996Z.Phys.A} and shown in Fig.~\ref{fig2}(b).
It is seen that the $\hbar\omega$ increases with spin.

Below the bandhead ($I<8\hbar$), $\hbar\omega$ is negative. This indicates the angular
momentum of the rotor anti-aligns along the spin, which is consistent
with the results of $\theta_{IR}$ shown in Fig.~\ref{fig1}(b).

At the bandhead, $\hbar\omega$ is near zero. The collective rotation is just
starting and rather small.

At $I=13\hbar$, a kink appears. As discussed in Ref.~\cite{Frauendorf1997NPA},
this is the evidence of the onset of the aplanar rotation. A dashed-dotted line
is plotted to label this position. Note that the spin region ($8\leq I<13\hbar$)
from the bandhead to the kink are usually called as chiral vibration region,
which in fact is a planar rotation~\cite{Frauendorf1997NPA, Olbratowski2004PRL,
Olbratowski2006PRC, Grodner2018PRL}.

At $I=15$-$17\hbar$, the spin region of the best degeneracy of the
doublets, the $\hbar\omega$ of the doublet bands are very similar.
This gives a hint that the angular momentum geometries of the doublets
are similar.

To examine the angular momentum coupling modes of the system, the normalized
orientation parameter $o$ is calculated~\cite{Starosta2001NPA, Starosta2002PRC}:
\begin{align}\label{eq6}
 o=\frac{\langle \mathcal{L}|\bm{R}\cdot(\bm{j}_p\times\bm{j}_n) |\mathcal{L}\rangle}{
 \sqrt{\langle\mathcal{L}|\bm{j}_p^2|\mathcal{L}\rangle}
 \sqrt{\langle\mathcal{L}|\bm{j}_n^2|\mathcal{L}\rangle}
 \sqrt{\langle\mathcal{L}|\bm{R}^2|\mathcal{L}\rangle}}~,
\end{align}
with $\langle \mathcal{L}|\bm{R}\cdot(\bm{j}_p\times\bm{j}_n) |\mathcal{L}\rangle
  =|\langle+|\bm{R}\cdot(\bm{j}_p\times\bm{j}_n) |-\rangle|$ and
$\langle\mathcal{L}|\bm{j}^2|\mathcal{L}\rangle
=\frac{1}{2}[\langle+|\bm{j}^2|+\rangle+\langle-|\bm{j}^2|-\rangle]$
($\bm{j}$ denotes $\bm{j}_p$, $\bm{j}_n$, and $\bm{R}$).
Here, $|+\rangle$ and $|-\rangle$ denote the wave functions of yrast
and yrare bands, and $|\mathcal{L}\rangle$ the wave function of
left-handed state in the intrinsic frame. In classical mechanics,
the normalized orientation parameter would vary between $o=1$ for
mutually perpendicular vectors and $o=0$ for planar
vectors~\cite{Starosta2001NPA, Starosta2002PRC}.

The result of the normalized orientation parameter was given for the
static chirality~\cite{Starosta2001NPA, Starosta2002PRC} or
for the bandhead~\cite{Grodner2018PRL}. Here we present it for
the whole spin region in Fig.~\ref{fig2}(d).

At $I=0$, $o=0$. This indicates a planar angular momentum geometry and
is consistent with the result that the effective angles $\theta_{pn}$,
$\theta_{Rp}$, and $\theta_{Rn}$ are about 120$^\circ$ (cf. Fig.~\ref{fig1}(a)).

With the increase of spin, $o$ first increases and then decreases,
corresponding to the appearance and disappearance of the aplanar rotation.
It shows strong correlation with the energy difference $\Delta E$
of the doublet bands (cf. Fig.~\ref{fig2}(c)). At $I=15$-$17\hbar$,
$o$ reaches to the maximal value, corresponding to the smallest
$\Delta E$ and the static chirality. It is also noted that
the maximal value of $o$ is not 1. This is consistent
with the result that the effective angles $\theta_{Rp}$,
$\theta_{Rn}$, and $\theta_{pn}$ are not $90^\circ$ at this spin region
as shown in Fig.~\ref{fig1}(a). Hence, one concludes that the angular
momenta of the rotor, the proton particle, and the neutron hole are not
ideally mutually perpendicular to each other at the static chiral
region. Nevertheless, the aplanar angular momentum
geometry at the static chiral region is better than that
near the bandhead.

The angular momentum geometry can also be illustrated by its profile on
the $(\theta,\varphi)$ plane $\mathcal{P}(\theta,\varphi)$, i.e., the
azimuthal plot~\cite{F.Q.Chen2017PRC}. Here, $(\theta,\varphi)$
are the tilted angles of the angular momentum with respect
to the intrinsic reference frame. In the calculations, we choose
1, 2, and 3 axes as short ($s$), long ($l$), and intermediate ($i$) axes,
respectively. Thus, $\theta$ is the angle between the angular momentum
and the $i$-axis, and $\varphi$ is the angle between the projection of the
angular momentum on the $s$-$l$ plane and the $s$-axis.

In Fig.~\ref{fig3}, the obtained profiles $\mathcal{P}(\theta,\varphi)$
are shown at $I=8$, 13, 15, and $20\hbar$ for the doublet bands.
It is observed that the maxima of the $\mathcal{P}(\theta,\varphi)$
always locate at $\varphi=45^\circ$ for all cases, since the angular
momentum has the same distributions along the $s$- and $l$- axes for
the current symmetric particle-hole configuration with
triaxial deformation $\gamma=30^\circ$. In addition, the $\mathcal{P}(\theta,\varphi)$
is symmetric with respect to the $\theta=90^\circ$ line. This is expected since the broken
chiral symmetry in the intrinsic reference frame has been fully restored in the PRM
wave functions. Hence, in the following, only the value of the $\theta$
($\leq 90^\circ$) is given when mentioning the position of the
maximal $\mathcal{P}(\theta,\varphi)$.

For the bandhead $I=8\hbar$, the angular momentum for yrast band mainly
orientates at $\theta=90^\circ$, namely, a planar rotation within the $s$-$l$
plane. The angular momentum for yrare band orientates at $\theta\sim 60^\circ$,
in accordance with the interpretation of chiral vibration along the $\theta$
direction (i.e., with respect to the $s$-$l$ plane).
For $I=13\hbar$, the angular momenta orientate at $\theta \sim 70^\circ$
for yrast band and $\theta \sim 50^\circ$ for yrare band. Starting from this
spin, the rotational mode of the yrast band changes from planar
to aplanar rotation. This is consistent with the appearance of kink in
the rotational frequency plot shown in Fig.~\ref{fig2}(b).
For $I=15\hbar$, the $\mathcal{P}(\theta,\varphi)$ of the yrast and yrare
bands are rather similar, which demonstrates the occurrence of static
chirality. The angular momenta orientate at $\theta\sim 45^\circ$ for both
bands. For $I=20\hbar$, the static chirality disappears. The angular momentum
for yrast band orientates to $\theta\sim 20^\circ$, while that for yrare
band to $\theta \sim 30^\circ$. The small values of $\theta$ correspond
to the fact that the angular momentum has large component along the $i$-axis.

Therefore, from the investigations of the azimuthal plots in Fig.~\ref{fig3},
we confirm that the rotational mode at bandhead is indeed a planar rotation. Then,
how to understand the results that the effective angles $\theta_{pn}$, $\theta_{Rp}$,
and $\theta_{Rn}$ are about $90^\circ$? We turn to investigate the vector lengths
of the angular momenta.

\begin{figure}[!ht]
  \begin{center}
    \includegraphics[width=7.5 cm]{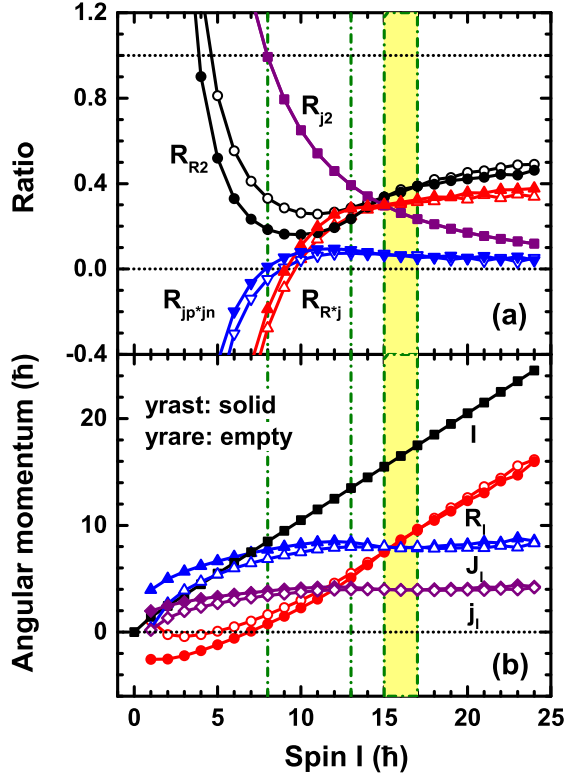}
    \caption{(a) Ratios $\langle \bm{R}^2\rangle/\langle \bm{I}^2 \rangle$,
    $\langle \bm{j}_p^2+\bm{j}_n^2 \rangle /\langle \bm{I}^2\rangle$,
    $\langle 2\bm{R}\cdot \bm{J}\rangle /\langle \bm{I}^2\rangle $,
    and $\langle 2\bm{j}_p\cdot\bm{j}_n\rangle/\langle \bm{I}^2 \rangle$
    (labeled as $\textrm{R}_{R^2}$, $\textrm{R}_{j^2}$,
    $\textrm{R}_{R*j}$, $\textrm{R}_{j_p*j_n}$, respectively)
    as functions of spin for the yrast and yrare bands.
    (b) Angular momentum vector projection
    along the total spin $I=\sqrt{\langle \bm{I}^2\rangle}$ of the rotor $R_{I}
    =\langle \bm{R}\cdot\bm{I}\rangle/\sqrt{\langle \bm{I}^2\rangle}$,
    the particles $J_I=\langle \bm{J}\cdot\bm{I}\rangle/\sqrt{\langle \bm{I}^2\rangle}$
    and $j_I=\langle \bm{j}_p\cdot\bm{I}\rangle/\sqrt{\langle \bm{I}^2\rangle}
    =\langle \bm{j}_n\cdot\bm{I}\rangle/\sqrt{\langle \bm{I}^2\rangle}$
    as functions of spin for the yrast and yrare bands.}\label{fig4}
  \end{center}
\end{figure}

The angular momenta of the rotor, the proton particle, and the
neutron hole are coupled to obtain the total spin as $\bm{I}=\bm{R}+\bm{J}$
with $\bm{J}=\bm{j}_p+\bm{j}_n$. As a consequence, $\bm{I}^2$ can be
decomposed as
\begin{align}
 \bm{I}^2=\bm{R}^2+(\bm{j}_p^2+\bm{j}_n^2)+2\bm{R}\cdot \bm{J}+2\bm{j}_p\cdot\bm{j}_n.
\end{align}
The ratios $\langle \bm{R}^2\rangle/\langle \bm{I}^2 \rangle$,
$\langle \bm{j}_p^2+\bm{j}_n^2 \rangle /\langle \bm{I}^2\rangle$,
$\langle 2\bm{R}\cdot \bm{J}\rangle /\langle \bm{I}^2\rangle $,
and $\langle 2\bm{j}_p\cdot\bm{j}_n\rangle/\langle \bm{I}^2 \rangle$
(labeled as $\textrm{R}_{R^2}$, $\textrm{R}_{j^2}$, $\textrm{R}_{R*j}$,
$\textrm{R}_{j_p*j_n}$, respectively) as functions of spin for the doublet bands
are calculated and shown in Fig.~\ref{fig4}(a). Obviously, the sum of these four
ratios are equal to 1.

From Fig.~\ref{fig4}(a), it is seen that $\textrm{R}_{j^2}$ decreases in a
hyperbola-like behavior, since $\langle \bm{j}_p^2+\bm{j}_n^2 \rangle=j_p(j_p+1)+j_n(j_n+1)$
is a constant in the single-$j$ shell model, while $\langle \bm{I}^2 \rangle=I(I+1)$ increases
in term of $I^2$. For the others, the $\textrm{R}_{R*j}$ increases gradually,
the $\textrm{R}_{j_p*j_n}$ first increases and then keeps nearly constant above the bandhead,
and the $\textrm{R}_{R^2}$ first decreases and then increases.

In detail, both the $\textrm{R}_{R*j}$ and the $\textrm{R}_{j_p*j_n}$
give negative contributions below the bandhead. At the bandhead,
the $\textrm{R}_{j_p*j_n}$ is zero, and above the bandhead, its contribution to
the total spin is rather small. For the $\textrm{R}_{j^2}$, its
contribution is much larger than 1 below the bandhead. At
the chiral vibration region ($8\leq I<13\hbar$), it still has a major contribution ($\geq$ 40\%)
to the total spin. At the static chiral region, its contribution is similar
as those of $\textrm{R}_{R^2}$ and $\textrm{R}_{R*j}$. However, beyond this region, it
becomes much smaller than $\textrm{R}_{R^2}$ and $\textrm{R}_{R*j}$. This is because
the angular momentum of the rotor plays more and more essential roles than those
of particle and hole as the spin increases. At the bandhead, the angular momentum
of the rotor is rather small in comparison with those of particle and hole. As a
result, although it is perpendicular to $\bm{j}_p$ and $\bm{j}_n$, it does not
indicate a aplanar rotation and good chirality. Bear this in mind, the total
angular momentum for the yrast band still lies in the $s$-$l$ plane (cf. Fig.~\ref{fig3}).

It is also noted that the $\textrm{R}_{R^2}$ of the doublet bands are quite different
at the chiral vibration region. This is attributed to that the angular momentum
of the rotor lies mainly in the $s$-$l$ plane for the yrast band,
while deviates from this plane for the yrare band in the chiral vibration region.
Such differences cause the energies of the doublet bands are different as
shown in Fig.~\ref{fig2}(c). It also provides additional
information that the static chirality is not realized yet.

From the above analysis, one concludes that the total spin below the static
chiral region ($I<15\hbar$) mainly comes from the proton and the neutron, in
the static chiral region ($15\leq I \leq 17\hbar$) also from the rotor,
and beyond the static chiral region ($I>17\hbar$) mainly
from the rotor. The paradox is caused by the fact that the angular momentum of the rotor
is much smaller than those of proton and neutron near the bandhead.
To show this more clearly, the projections of the rotor $R_I$
and the particles $J_I$ and $j_I$ along the total spin are calculated:
\begin{align}
 R_I&=\langle \bm{R}\cdot\bm{I}\rangle/\sqrt{\langle \bm{I}^2\rangle}, \\
 J_I&=\langle \bm{J}\cdot\bm{I}\rangle/\sqrt{\langle \bm{I}^2\rangle},\\
 j_I&=\langle \bm{j}_p\cdot\bm{I}\rangle/\sqrt{\langle \bm{I}^2\rangle}
    =\langle \bm{j}_n\cdot\bm{I}\rangle/\sqrt{\langle \bm{I}^2\rangle}.
\end{align}
Note that here $J_I=2j_I$ and $R_I+J_I=\sqrt{\langle \bm{I}^2\rangle}
=\sqrt{I(I+1)}$. The obtained results are given in Fig.~\ref{fig4}(b).

With the increase of spin, $R_I$ increases gradually. The $J_I$ as
well as the $j_I$ increase slightly below the kink ($I\leq 13\hbar$),
and keep nearly constant in the above ($I>13\hbar$). Below the bandhead, $R_I$
contributes negatively as it anti-aligns along the total
spin. At the bandhead, it is very small. Then it becomes
gradually comparable with $j_I$, but is still smaller than
$J_I$. At static chiral region, $R_I\approx J_I$. This is consistent
with the result that the value of the $\theta$ for the maximal
$\mathcal{P}(\theta,\varphi)$ is about $45^\circ$ (cf. Fig.~\ref{fig3}).
In this case, the energy difference between the doublet
bands is the smallest. Beyond the static chiral region, $R_I$ becomes
larger than $J_I$ and responsible for the increase of total spin, which
results in a principal axis rotation along the $i$-axis.
Therefore, with the increase of spin, the angular momentum of
the rotor plays gradually more and more important roles than
those of proton particle and neutron hole.

In summary, the paradox on the previous interpretation for the nuclear
chiral geometry based on the effective angle has been clarified by
reexamining the system with the particle-hole
configuration $\pi (1h_{11/2})^1 \otimes \nu(1h_{11/2})^{-1}$
and rotor with deformation parameter $\gamma=30^\circ$.
According to the studies of normalized orientation parameter of the system
and the azimuthal plot of the total angular momentum, we confirm that chiral
rotation does indeed exist only at a certain high spin region. Further
study for the angular momentum shows that the paradox is caused by the
fact that the angular momentum of the rotor is much smaller than
those of the proton and the neutron near the bandhead.
Hence, it does not support a chiral rotation interpretation
near the bandhead. The nuclear chiral geometry based on the
effective angle makes sense only when the angular momentum
of the rotor becomes comparable with those of the proton
and the neutron at the certain spin region.

The authors would like to thank S. Q. Zhang and P. W. Zhao for fruitful
discussions. Financial support for this work was provided in parts by Deutsche
Forschungsgemeinschaft (DFG) and National Natural Science Foundation
of China (NSFC) through funds provided to the Sino-German CRC 110
``Symmetries and the Emergence of Structure in QCD'',
the Major State 973 Program of China No.~2013CB834400, and the
NSFC under Grants No.~11335002 and No.~11621131001.

\end{CJK}

\end{document}